\documentstyle[seceq,twoside]{ptptex}
\notypesetlogo  
\title{Covariant Quark-Representation of Composite Meson Systems\\ 
and Chiral Symmetry   }
\author{%
Shin {\sc Ishida}, Muneyuki {\sc Ishida}$^{*}$ and Tomohito {\sc Maeda} }
\inst{%
 Atomic Energy Research Institute, 
College of Science and Technology\\
Nihon University, Tokyo 101-0062, Japan\\
$^{*}$Department of Physics, Tokyo Institute of Technology\\
Tokyo 152-8551, Japan
}
\recdate{%
\today
}
\abst{%
Assuming the spin-independence for confining force, we give a covariant
quark representation of general composite meson systems with definite 
Lorentz transformation properties. For benefit of this representation
we are able to deduce automatically the transformation rules of composite
mesons for general symmetry operations from those of constituent (exciton)
quarks. Applying this we investigate especially physical implication 
of chiral symmetry for the meson systems, and point out a possibility of 
existence of new meson multiplets.
}

\begin{document}
\maketitle

\setcounter{tocdepth}{4}

\section{Introduction}

There are the two contrasting view points of composite quark-antiquark
mesons: The one is non-relativistic, based on the approximate symmetry of
$LS$-coupling in the non-relativistic quark model (NRQM); 
while the other is relativistic, based on the 
dynamically broken chiral symmetry 
typically displayed in the Nambu Jona-Lasinio (NJL) model. 
The $\pi$-meson (or $\pi$-nonet) is now widely believed to have a
dual nature of non-relativistic particle with $(L,S)=(0,0)$
and also of relativistic particle as a Nambu-Goldston boson with
$J^P=0^-$ in the case of spontaneous breaking of chiral symmetry. However,
no successful attempts to unify the above two view points have been yet proposed.
On the other hand we have developed the 
covariant oscillator quark model (COQM)\cite{rf1,rf3,rf2,rfmass,rf5}
for many years as a covariant extension of NRQM, which is based on 
the boosted $LS$-coupling scheme. The meson wave functions (WF)
in COQM are tensors in the $\tilde U(4)\bigotimes O(3,1)$ space
and reduce at the rest frame to those in the 
$SU(2)_{\rm spin}\bigotimes O(3)_{\rm orbit}$ space in NRQM.
The COQM has been successful especially in treating the
$Q\bar Q$ meson system and the 
($q,\bar Q$) or ($Q,\bar q$) meson system, leading, respectively, 
to a satisfactory understanding of radiative transitions 
and to the same
weak form factor relations as in the heavy quark effective theory (HQET).
However, in COQM no 
consideration on chiral symmetry has been given and it is not able to 
explain the dual nature of $\pi$ meson.

The purpose of this paper is to get rid of this defect in COQM and 
is to give a unified view point of the two contrasting ones
of the composite meson systems,
extending the tensors of WF from the restricted ones necessary only in the boosted $LS$
coupling scheme to the general ones in the  $\tilde U(4)\bigotimes O(3,1)$ space,
which are required for taking into account chiral symmetry.

\section{Covariant Framework for Describing Composite Mesons}
For meson WF described by $\Phi_A{}^B(x_1,x_2)$ ($x_1,x_2$ denoting the 
space-time coordinate and 
$A=(\alpha ,a)(B=(\beta ,b))$ denoting the Dirac spinor and 
flavor indices  of constituent quark (anti-quark)) 
we set up the bilocal Yukawa equation\cite{rf2}
\begin{eqnarray}
\left[ \frac{\partial^2}{\partial X_\mu^2}
 - {\cal M}^2(x_\mu ,\frac{\partial}{\partial x_\mu} )\right]
\Phi_A{}^B(X,x)=0
\label{eq1}
\end{eqnarray}
($X(x)$ denoting the center of mass (CM) (relative) coordinate of meson),
where the  ${\cal M}^2$ is squared mass operator including
only a central, Dirac-spinor-independent\footnote{
In the boosted $LS$-coupling scheme the squared mass operator ${\cal M}^2$
was assumed to be only Pauli-spinor independent, while in the present scheme 
it is assumed more generally to be Dirac-spinor independent. 
} 
confining potential.
The WF is separated into the plane wave describing 
CM motion and the (Fierz-component) internal WF as 
\begin{eqnarray}
\Phi_A{}^B(x_1,x_2)=\sum_{{\bf P}_n,n}(e^{iP_nX}\Psi_{n,A}{}^{(+)B}(x,P_n)
+e^{-iP_nX}\Psi_{n,A}{}^{(-)B}(x,P_n)),
\label{eq2}
\end{eqnarray}
where the Fierz components $\Psi_n^{(\pm )}$ are eigenfunctions of ${\cal M}^2$ as 
\begin{eqnarray}
{\cal M}^2(x_\mu ,\frac{\partial}{\partial x_\mu},P_n )\ \Psi_n^{(\pm )}
  = M_n^2\  \Psi_n^{(\pm )}\ ,   
\label{eq2c}
\end{eqnarray}
$P_{n,\mu}^2=-M_n^2,\ P_{n,0}=\sqrt{M_n^2+{\bf P}_n^2}$;
and the label 
$(\pm )$ represents the positive (negative) frequency part;
and $n$ does a freedom of excitation. We have the following field 
theoretical expression for the WF in mind 
as a guide for developing the present semi-phenomenological approach:
\begin{eqnarray}
\Phi_A{}^B(x_1,x_2) &=& \sum_n[\langle 0|\psi_A(x_1)\bar\psi^B(x_2)|M_n\rangle
+\langle M_n^c|\psi_A(x_1)\bar\psi^B(x_2)|0\rangle ] ,
\label{eq3}
\end{eqnarray}
where $\psi_A(\bar\psi^B)$ denotes the quark field (its Pauli-conjugate)
and $|M_n\rangle\ (\langle M_n^c|)$ does the composite meson 
(its charge conjugate) state,
and the first (second) term in the RHS corresponds to the positive (negative) frequency part
in Eq.~(\ref{eq2}).
The internal WF is, concerning the Dirac-spinor-dependence, expanded in terms of 
a complete set $\{ W_i \}$ of free bi-Dirac spinors of 
quarks and anti-quarks; and 
the internal WF is expressed as
\begin{eqnarray}
\Psi_A^{(\pm )B}(x,P_n)=\sum_{i}W_{i\alpha}^{(\pm )\beta}(P_n)
\phi_a^{(\pm )b}(x,P_n),\nonumber \\ 
\phi^{(\pm )}(x,P_n)=\epsilon_i\langle \bar W_i^{(\mp )}\Psi^{(\pm)}
\rangle ,
\label{eq4}
\end{eqnarray}
where $\langle A\rangle$ means trace of $A$. 
The ortho-normal
relations $\langle \bar W_i^{(\mp)}W_j^{(\pm )} \rangle 
=\epsilon_{i}\delta_{ij}$ holds for the Pauli-conjugate of WF, defined by
$\bar W_i^{(\mp )} \equiv \gamma_4 W_i^{(\pm )\dagger} \gamma_4$,
where $\bar W_i^{(\mp)}$ is related with the $W_i^{\mp}$, and 
the $\epsilon_{i}$ and $\delta_{ij}$ denote,respectively, the sign
and the Kronecker symbols, see the appendix A).


\section{Complete Set of Spin Wave Function\\
and Composite Mesons with Definite Spin}
We set up the conventional ``free'' Dirac spinors with four-momentum of 
composite meson itself $P=P_M$,
$D_{q,\alpha}(P)\equiv (u_{q,\alpha}(P,s_q),v_{q,\alpha}(P,s_q)$
($s_q=\pm$ representing the spin up-down)) for quarks and  
$\bar D_{\bar q}{}^\beta (P)\equiv  (\bar v_{\bar q}^{\ \ \beta}(P,s_{\bar q}),
\bar u_{\bar q}^{\ \ \beta}(P,s_{\bar q})$($s_{\bar q}=\pm$ representing spin 
up-down)) for anti-quarks.
It is to be noted that all four spinors for both ``quarks 
and anti-quarks"  are necessary\footnote{
For understanding this it may be useful to take an analogy of the Bethe-Salpeter
amplitude of deuteron. In expanding the amplitude all $4\times 4$ members of
direct product of both Dirac spinors for constituent proton and neutron are necessary.
}
 to describe the spin WF of mesons.
Then the complete set of bi-Dirac spinors is 
given by\footnote{
In the following from \S 3 to \S 4.3 we give only the expressions of 
$(+)$-frequency parts,
and consider only the ground states of composite system, disregarding the 
relative coordinates. In Eq.(\ref{eq9}) considerations on the freedom of Pauli-spin
are neglected. We have given detailed considerations on this problem and useful formulas
in Appendix A.
}
\begin{eqnarray}
\{ W^{(+)}(P) \} &:& \nonumber\\
U(P) &=& u_q(p_1,s_q)\bar v_{\bar q}(p_2,s_{\bar q})|_{p_{i,\mu}=\kappa_iP_\mu}
=u_+({\mib P},s_q)\bar v_+({\mib P},s_{\bar q}),\nonumber\\
C(P) &=& u_q(p_1,s_q)\bar u_{\bar q}(p_2,s_{\bar q})|_{p_{i,\mu}=\kappa_iP_\mu}
=u_+({\mib P},s_q)\bar v_-(-{\mib P},s_{\bar q}),\nonumber\\
D(P) &=& v_q(p_1,s_q)\bar v_{\bar q}(p_2,s_{\bar q})|_{p_{i,\mu}=\kappa_iP_\mu}
=u_-(-{\mib P},s_q)\bar v_+({\mib P},s_{\bar q}),\nonumber\\
V(P) &=& v_q(p_1,s_q)\bar u_{\bar q}(p_2,s_{\bar q})|_{p_{i,\mu}=\kappa_iP_\mu}
=u_-(-{\mib P},s_q)\bar v_-(-{\mib P},s_{\bar q}),
\label{eq9}
\end{eqnarray}
where
$u_+({\mib P})(\bar v_+({\mib P}))$ and 
$u_-(-{\mib P})(\bar v_-(-{\mib P}))$ denote the Dirac spinors with 
positive energy and momentum ${\mib P}$ and with negative energy 
and momentum $-{\mib P}$, respectively, describing quarks 
(anti-quarks). 
These energy and momentum concern with the total meson, while in Eq.(\ref{eq9}) 
we have defined technically the momenta of
``constituent quarks"\footnote{
In so far as concerned with Eqs.~(\ref{eq9}) and (\ref{eq10}) the quantities $\kappa_i$
and accordingly $m_i$ are arbitrary and have no physical meaning. However, $m_i$
have proved to be the effective masses of constituent quarks through the phenomenological
applications\cite{rfmass,rf5} of COQM so far made.
} as 
\begin{eqnarray}
p_{i,\mu} &\equiv&  \kappa_iP_\mu ,\ p_{i,\mu}^2=-m_i^2;\ P_\mu^2=-M^2,
M=m_1+m_2\nonumber\\
 (\kappa_{1,2} &\equiv& m_{1,2}/(m_1+m_2); \ \ \kappa_1+\kappa_2=1).
\label{eq10}
\end{eqnarray}
The respective members in Eq.(\ref{eq9})
satisfy a couple of the corresponding free Dirac equations in momentum 
space (which are equivalent to the (conventional or 
new-type of) Bargman-Wigner Equations) and are expressed 
in terms of their irreducible composite meson WF 
as follows:
\begin{eqnarray}
{\rm (Non\ Rela.\  comp.)} & & \nonumber\\
 (iP\gamma^{(1)} &+& M)U(P)=0,\ U(P)(-iP\gamma^{(2)}+M)=0;\nonumber\\
  U_A{}^B(P) &=& \frac{1}{2\sqrt{2}}
   [(i\gamma_5P_{s,a}^{(NR)b}(P)+i\gamma_\mu V_{\mu ,a}^{(NR)b}(P))
    (1+\frac{iP\cdot\gamma}{M}) ]_\alpha{}^\beta ,\nonumber \\
{\rm (Semi\ Rela.\ comp.)} & & \nonumber\\
  \bar q{\rm -type}\ \ \ \ (iP\cdot\gamma^{(1)} &+& M)C(P)=0,\ 
                      C(P)(iP\cdot\gamma^{(2)}+M)=0;\nonumber\\
  C_A{}^B(P) &=& \frac{1}{2\sqrt{2}}
   [(S_{a}^{(\bar q) b}(P)+i\gamma_5\gamma_\mu A_{\mu ,a}^{(\bar q)b}(P))
    (1-\frac{iP\cdot\gamma}{M}) ]_\alpha{}^\beta ,\nonumber\\
  q{\rm -type}\ \ \ \ (-iP\gamma^{(1)} &+& M)D(P)=0,\ 
                      D(P)(-iP\gamma^{(2)}+M)=0;\nonumber\\
  D_A{}^B(P) &=& \frac{1}{2\sqrt{2}}
   [(S_{a}^{(q) b}(P)+i\gamma_5\gamma_\mu A_{\mu ,a}^{(q)b}(P))
    (1+\frac{iP\cdot\gamma}{M}) ]_\alpha{}^\beta ,\nonumber\\
{\rm (Extrly.Rela.\ comp.)} & & \nonumber\\
  (-iP\gamma^{(1)} &+& M)V(P)=0,\ V(P)(iP\gamma^{(2)}+M)=0;\nonumber\\
  V_A{}^B(P) &=& \frac{1}{2\sqrt{2}}
   [(i\gamma_5P_{s,a}^{(ER)b}(P)+i\gamma_\mu V_{\mu ,a}^{(ER)b}(P))
    (1-\frac{iP\gamma}{M}) ]_\alpha{}^\beta ,\ \ \ \ \ \ \ \ \ \ \ 
\label{eq11}
\end{eqnarray}
where  
all vector and axial-vector mesons satisfy the Lorentz 
conditions, $P_\mu V_\mu (P)=P_\mu A_\mu (P)=0$.
Here it is to be noted that, in each type of the above members, 
the number of freedom counted both in the quark representation and 
in the meson representation is equal,
as it should be ($2\times 2=4$ and $1+3=4$, respectively).
Also it may be amusing to note that each constituent quarks 
in all the above members is in 
``parton-like motion,''
having the same 3-dimentional velocity as that of total mesons.
(For example, in $V(P),\ {\mib v}_{1,2}=
\frac{{\mib p}^{(1,2)}}{p_0^{(1,2)}}=
\frac{-\kappa_{1,2}{\mib P}_M}{-\kappa_{1,2}P_{M,0}}=
{\mib v}_M$.)

\section{Transformation Properties of Composite Meson Systems\\
 and Chiral Symmetry}
For benefit of covariant quark-representation of mesons given in \S 3 we can
deduce automatically the transformation rules of composite mesons for general
symmetry operations as follows:

\subsection{Charge conjugation}
Charge conjugation properties of the bi-spinors 
and, correspondingly, of the composite mesons are derived 
from those of quarks as follows:
\begin{eqnarray}
{\rm quark\ field:} &\ &  \psi_\alpha (x)\rightarrow 
  \psi_\alpha^c(x)=U_C^{-1}\psi_\alpha (x)U_C
       =(C^{-1})_{\alpha\alpha '}\bar\psi^{\alpha '}(x)\nonumber\\
  &\ & \bar\psi^\beta (x)\rightarrow 
  \overline{\psi^c}^\beta (x)=U_C^{-1}\bar\psi^\beta (x)U_C
       =C^{\beta\beta '}\psi_{\beta '}(x) \\
  &\ & (CC^\dagger =1,\ C=\gamma_4\gamma_2,\ 
        C\gamma_\mu C^{-1}=-^t\gamma_\mu ). \nonumber \\
{\rm Internal\ meson} &{\rm WF:}&  \nonumber\\
    \Psi_A^{(+)B} (P,x) 
 ( &\approx& \langle 0|\psi_A(x_1)\bar\psi^B(x_2)|M\rangle )
        \ \ \ \ \ \ \ \ \ \ \ \ \ \nonumber\\
   &\rightarrow& \Psi_A^{c,(+)B} (P,x)  
 ( \approx \langle 0|\psi_A(x_1)\bar\psi^B(x_2)|M^c\rangle \nonumber\\
   &=& \langle 0|\  U_C^{-1} \psi_A(x_1)U_C\ U_C^{-1}\bar\psi^B(x_2)U_C|M\rangle
              \nonumber\\
   &=& \langle 0|\  (C^{-1})_{AA'} \bar\psi^{A'}(x_1) C^{BB'}\psi_{B'}(x_2)|M\rangle
              \nonumber\\
   &=& (C^{-1})_{AA'} (-1)\langle 0|\   \psi_{B'}(x_2) \bar\psi^{A'}(x_1) |M\rangle
          (^tC)^{B'B}   ) \nonumber\\
   &=& (C^{-1})_{AA'} {}^t\Psi^{(+)}(P,-x)^{A'}{}_{B'}C^{B'B},
                \nonumber \\
{\rm Spinor}  &{\rm WF:}& \nonumber\\
             && U_{P_s}^{(NR)} \leftrightarrow U_{P_s}^{(NR)},\ \  
                   U_{V_\mu}^{(NR)} \leftrightarrow -U_{V_\mu}^{(NR)}, \nonumber\\
             && C_S^{\bar q} \leftrightarrow D_S^q,\ \ 
                   C_{A_\mu}^{\bar q} \leftrightarrow D_{A_\mu}^q,\nonumber\\
             && U_{P_s}^{(ER)} \leftrightarrow U_{P_s}^{(ER)},\ \  
                   U_{V_\mu}^{(ER)} \leftrightarrow -U_{V_\mu}^{(ER)}, \nonumber\\
{\rm Composite\ } &{\rm meson}& {\rm \ WF:}  \nonumber\\
      P_{s,a}^{(NR)b} &\leftrightarrow& P_{s,b}^{(NR)a},\ \    
      V_{\mu ,a}^{(NR)b} \leftrightarrow -V_{\mu ,b}^{(NR)a},\nonumber\\
      S_{a}^{(\bar q)b} &\leftrightarrow& S_{b}^{(q)a},\ \ 
      A_{\mu ,a}^{(\bar q)b} \leftrightarrow A_{\mu ,b}^{(q)a},\nonumber\\
      P_{s,a}^{(ER)b} &\leftrightarrow& P_{s,b}^{(ER)a},\ \ 
      V_{\mu ,a}^{(ER)b} \leftrightarrow -V_{\mu ,b}^{(ER)a}.
\label{eq12}
\end{eqnarray}

\subsection{Chiral transformation}
Chiral transformation properties of composite mesons are 
also derived straightforwardly from those of the bi-spinors.
For example, for $SU(3)$ chiral transformation   
\begin{eqnarray}
\Psi_A{}^B(P,x) &\rightarrow & 
   [e^{i\alpha^i\frac{\lambda^i}{2}\gamma_5}\Psi (P,x)
    e^{i\alpha^i\frac{\lambda^i}{2}\gamma_5}]_A{}^B.
\label{eq18}
\end{eqnarray}
leading to the results:
\begin{eqnarray}
[M^{(J)}(P)]'_i  &=& [^tS^{(J)}(P,\alpha )]_{ij}[M^{(J)}(P)]_j
          \ \ \ \ \ J=(0,1) \nonumber \\  
M^{(0)}_i
  &\equiv & {}^t [P_s^{(NR)},S^{(R\bar q)}, S^{(Rq)},
  P_s^{(ER)}]_i   , \nonumber\\ 
M^{(1)}_i
  &\equiv & {}^t[V_\mu^{(NR)},A_\mu^{(R\bar q)}, A_\mu^{(Rq)},
  V_\mu^{(ER)}]_i ,
\label{eq19}
\end{eqnarray}
where $[S^{(J)}]$ denotes a unitary matrix, 
of which explicit form is omitted here. (For infinitesimal transformation, 
see appendix B).

\subsection{Light-quark meson system---``chiral SU(6) multiplet''}
The quark representation applying to the 
light-quark mesons is obtained by the linear transformation of 
the bi-Dirac spinors given in \S 3 as follows:
\begin{eqnarray}
U_{P_s,\alpha}^{(N,E)\beta} &\equiv & \frac{1}{\sqrt{2}}\ 
   (U_{P_s}\pm V_{P_s})_\alpha{}^\beta 
   = \frac{1}{2}\ [( i \gamma_5,-\gamma_5 v\cdot\gamma )]_\alpha{}^\beta ;\ \ \ 
        P_s^{(N,E)}; C=(+,+) \nonumber\\
C_{S,\alpha}^{(N,E)\beta} &\equiv & \frac{(1,i)}{\sqrt{2}}\ 
   (D_S\pm C_S)_\alpha{}^\beta 
   = \frac{1}{2}\ [(1,-v\cdot\gamma )]_\alpha{}^\beta ;\ \ \ 
        S^{(N,E)}; C=(+,-) \nonumber\\
U_{V,\alpha}^{(N,E)\beta} &\equiv & \frac{1}{\sqrt{2}}\ 
   (U_{V}\pm V_{V})_\alpha{}^\beta 
   = \frac{1}{2}\ [(i\tilde\gamma_\mu ,-i\sigma_{\mu\nu}v_\nu )]_\alpha
    {}^\beta ;\ \ \   V^{(N,E)}; C=(-,-) \nonumber\\
C_{A,\alpha}^{(N,E)\beta} &\equiv & \frac{(1,i)}{\sqrt{2}}\ 
   (D_A\pm C_A)_\alpha{}^\beta 
   = \frac{1}{2}\ [(i\gamma_5 \tilde\gamma_\mu ,\gamma_5 \sigma_{\mu\nu}v_\nu )
   ]_\alpha{}^\beta ;\ \ \   A^{(N,E)}; C=(+,-)  \nonumber\\
\ \ \ \   & \ \  &\ \ \ \ 
\label{eq20}
\end{eqnarray}
($v_\mu\equiv P_\mu /M,\ \tilde\gamma_\mu v_\mu\equiv 0$; and
 $U_{P_s}$ denotes the coefficient bi-spinors of $P_s$ and so on ),
where we have given also the charge-conjugation parity of the 
corresponding (hidden flavor) composite mesons. 
The chiral transformation properties
 of the new bi-spinors 
are easily seen to be similar as the conventional ones as
\begin{eqnarray}
 &\  & 1 \leftrightarrow  i\gamma_5,\ \ 
-v\cdot\gamma\leftrightarrow -\gamma_5v\cdot\gamma ,\nonumber\\ 
 &\  & i\gamma_5\tilde\gamma_\mu 
  \leftrightarrow  i\tilde\gamma_\mu ,\ \ 
-i\sigma_{\mu\nu}v_\nu\leftrightarrow \gamma_5
\sigma_{\mu\nu}v_\nu .
\label{eq21}
\end{eqnarray}

\subsection{``Local chiral SU(6) field''}
Extending our considerations to include the $(-)$-frequency part,
we are led to a unified expression of what to be called,
Local Chiral SU(6) field, as
\begin{eqnarray}
\Psi_A{}^B(X) &=& \Psi_A^{(N)B}(X) + \Psi_A^{(E)B}(X) \nonumber\\
\Psi_A^{(N)B} &=& \frac{1}{2}\ \biggr[
 i\gamma_{5\alpha}{}^\beta P_{s,a}^{(N)b}
 +i\tilde\gamma_{\mu ,\alpha}{}^\beta V_{\mu ,a}^{(N)b}
 +1_\alpha{}^\beta S_{a}^{(N)b}
 +(i\gamma_5\tilde\gamma_\mu )_\alpha{}^\beta A_{\mu ,a}^{(N)b} \biggr]
 \nonumber\\
\Psi_A^{(E)B} &=& \frac{1}{2}\ \biggr[
(i\gamma_5\gamma_\mu )_\alpha{}^\beta \frac{\partial_\mu}{\sqrt{\partial^2}} 
  P_{s,a}^{(E)b}
  +(\sigma_{\mu\nu})_{\alpha}{}^\beta  \frac{\partial_\mu}{\sqrt{\partial^2}}
  V_{\nu ,a}^{(E)b} \nonumber \\
 & & \ \ \ \ \ \ \  +  
i\gamma_{\mu ,\alpha}{}^\beta \frac{\partial_\mu}{\sqrt{\partial^2}} S_{a}^{(E)b}
    +(i\gamma_5\sigma_{\mu\nu})_{\alpha}{}^\beta 
    \frac{\partial_\mu}{\sqrt{\partial^2}} A_{\nu ,a}^{(E)b} \biggr].
\label{eq22}
\end{eqnarray}

\section{Covariant Classification and Spectroscopy 
of Mesons and Chiral Symmetry}
\subsection{Level classification of ground states}
In the previous sections \S 2 and \S 3 we have presented a general covariant kinematical
framework for describing the (ground states of) composite meson systems with a definite
total quark spin. However, what kinds of mesons do really exist or not, that is, the meson
spectroscopy, is a dynamical (still unsolved) problem of QCD.

For this problem, it is useful to apply in our scheme 
a physical consideration of dynamically
broken chiral symmetry of QCD, typically displayed in the NJL model:
In the (ground state of) heavy quarkonium ($Q\bar Q$) system both quarks($Q$) and 
antiquarks($\bar Q$) are possible to do, since $m_Q > \Lambda_{\rm conf}$, 
only non-relativistic motions with positive energy,
and the non-relativistic $LS$-symmetry is good.
Accordingly the bi-spinor $U$ is considered to be applied to $Q\bar Q$
 system as a covariant spin WF.
In the (ground state of) heavy-light quark meson $Q\bar q$($q\bar Q$) system
the anti-quarks(quarks) make, since   $m_q \ll \Lambda_{\rm conf}$, 
relativistic motions both with positive and negative energies,
and the relativistic chiral symmetry concerning light antiquarks (quarks) 
is good.
Accordingly both the bi-spinors $U$ and $C$ ($U$ and $D$)
are to be applied to the  $Q\bar q$($q\bar Q$) system, and in this system
there is a possibility of 
existence of new composite scalar and axial-vector mesons(see
Eq.(\ref{eq11})).
In the (ground state of) light quark $q\bar q$-meson system 
both quarks $q$ and anti-quarks $\bar q$
make, since $m_q\ll\Lambda_{\rm conf}$, relativistic motions 
with both positive and negative energies, and chiral symmetry is good.
Accordingly the linear combinations (specified in \S 4.3) of
bispinors $U$ and $V$ are applied to the $q\bar q$-system, and
in this system there is a possibility of existence of an extra(, in 
addition, to a normal) set of 
composite pseudo-scalar and vector mesons.
Furthermore, the linear combinations of $C$ and $D$ are also applied, and normal and extra
sets of composite scalar and axial-vector mesons possibly exist as relativistic $S$-wave 
bound states. 

\begin{table}[t]
\caption{Level structure of the ground states of 
general quark meson systems } 
\begin{center}
\begin{tabular}{|l|l|c|l|l|}
\hline\hline
[Heavy quark system] & & Mass & Spin WF& Type of Mesons\\
\hline
\ \ $m_q<\Lambda_{\rm conf}$: $q$ Rela.& $q\bar Q$
            & $m_q+m_{\bar Q}$ & $u_+\bar v_+ ;\  u_-\bar v_+$ & $P_s,V_\mu ;S, A_\mu$ \\
\ \ \ \ ($q$ chiral sym. good) & & & & \\
\ \ $m_{\bar q}<\Lambda_{\rm conf}$: $\bar q$ Rela.& $Q\bar q$
            & $m_Q+m_{\bar q}$ & $u_+\bar v_+ ;\  u_+\bar v_-$ & $P_s,V_\mu ;S, A_\mu$ \\
\ \ \ \ ($\bar q$ chiral sym. good) & & & & \\
\ \ $m_Q,m_{\bar Q}>\Lambda_{\rm conf}$:Non-R& $Q\bar Q$
            & $m_Q+m_{\bar Q}$ & $u_+\bar v_+$ & $P_s,V_\mu$ \\
\ \ \ \ ($LS$ sym. good) & & & & \\
\hline
[Light quark system] & & & & \\
\hline
\ \ $m_q,m_{\bar q}<\Lambda_{\rm conf}$:Rela.& $q\bar q$
            & $m_q+m_{\bar q}$ & $\frac{1}{\sqrt 2}(u_+\bar v_+ \pm u_-\bar v_-)$ &
                       $P_s^{(N)},V_\mu^{(N)} ;P_s^{(E)},V_\mu^{(E)} $ \\
\ \ \ \ (chiral sym. good) &    &  &  $\frac{(1,-i)}{\sqrt 2}(u_+\bar v_- \pm u_-\bar v_+)$ &
                       $S^{(N)},A_\mu^{(N)} ;S^{(E)},A_\mu^{(E)} $ \\
\hline\hline
\end{tabular}
\end{center}
\end{table}

In the above discussion on the validity of chiral symmetry, 
we have supposed that $\Lambda_{\rm conf}\sim 1$GeV 
regardless of quark-flavor. 
Here it may be useful to note that the positive (negative) energy 
Dirac spinors with momentum ${\mib P}$ for quarks and 
antiquarks change, respectively, into negative (positive) 
energy ones with momentum -${\mib P}$ under the operation of $\gamma_5$ 
as 
\begin{eqnarray}
\gamma_5u_+({\mib P}) &=& u_-(-{\mib P}),\ \ 
\bar v_+({\mib P})\gamma_5=\bar v_-(-{\mib P}) .
\end{eqnarray}
The above expected level structure of the ground states of 
general light-through-heavy quark
meson systems is summarized in Table I.
 
\subsection{Level classification of excited states}
In the last subsection we have stated on our level classification scheme, focusing on the
ground-state mesons. In classifying the excited-state mesons we can proceed essentially
similarly\cite{rf1} as the case of ground state mesons. 
In (the present extended version of) COQM
in the pure-confining force limit, the masses of N-th excited states
are given by the formula 
$M_N^2=M_G^2+N\Omega\ (M_G\equiv M_0,\ \Omega$ being the inverse Regge slope), 
and their covariant spin wave functions
are defined by the same formulas as given in \S 3 
with substitution of constituent 
exciton-quark mass $m_i$ by 
\begin{eqnarray}
m_i^* &=& \gamma_N\ m_i\ \ (\gamma_N\equiv M_N/M_G) . \label{eq51}
\end{eqnarray}
The value of confinement momentum $\Lambda_{\rm conf}$ is 
conventionally considered to be 
\begin{eqnarray}
\Lambda_{\rm conf}  &\sim & 1 {\rm GeV}. \label{eq52}
\end{eqnarray}
The value of 
 $m_i^*$ for respective quark-configuration meson systems 
obtained by the formula (\ref{eq51}) 
are given in Table II. 
From this table we see that $m^* \ll \Lambda_{\rm conf}$ 
for the lower levels, especially for the ground and first-excited states 
in the light-quark meson systems, and accordingly we can expect 
that chiral symmetry for these states may be still good. 
Similarly,  in the light/heavy quark meson systems,
the chiral symmetry concerning the light quark 
is also good for the several lower levels.

\begin{table}[t]
\caption{The values of exciton-quark mass $m_i^*$ for $q\bar q$ and $q\bar Q$
meson systems. In the case of  $m_i^* < \Lambda_{\rm conf}\simeq 1$ GeV we can expect
that the chiral symmetry for the corresponding meson system is still effective.
The values of parameter $\beta$, representing the inverse of space-time extension of
the system, and of the oscillator quanta $\Omega$ are also given for reference.
The $\Omega$ is related with $\beta$ by 
$\Omega =(2(m_1+m_2)^2/(m_1m_2))\beta $.
The values\cite{rfmass} of constituent quark masses are taken as $m_n=0.375$,
$m_s=0.51$,  $m_c=1.70$ and $m_b=5.00$ (in GeV). }
\begin{center}
\begin{tabular}{|c|c|c|c|c|c|c|c|}
\hline
 & $n\bar n$ & $n\bar s$ & $s\bar s$ & $n\bar c$ & $s\bar c$ & $n\bar b$ & $s\bar b$ \\
\hline
$\Omega$/GeV$^2$ &1.09&1.14&1.13&1.96&1.70&4.56&3.71\\
$\beta$/GeV$^2$ &0.137&0.139&0.142&0.145&0.151&0.148&0.156\\ 
\hline
$(m_{q_1}^*,m_{\bar q_2}^*)$ &$m_{n}^*$ &$(m_{n}^*,m_{\bar s}^*)$ &
$m_s^*$ &$(m_{n}^*,m_{\bar c}^*)$ &$(m_{s}^*,m_{\bar c}^*)$ &
$(m_{n}^*,m_{\bar b}^*)$ & $(m_{s}^*,m_{\bar b}^*)$ \\
\hline
($L=0$) &0.375&(0.375,0.51)&0.51&(0.375,1.7)&(0.51,1.7)&(0.375,5)&(0.51,5)\\
($L=1$) &0.64&(0.59,0.80)& 0.74 &(0.45,2.1)&(0.59,2.0)&(0.40,5.4)&(0.54,5.3)\\
($L=2$) &0.83&(0.74,1.0)& 0.9 &(0.52,2.3)&(0.66,2.2)&(0.43,5.7)&(0.57,5.6)\\
\hline
\end{tabular}
\end{center}
\end{table}

The quantum numbers $P$ and $C$ are given for the respective 
orbitally $L$-th excited 
light-quark mesons as follows:
\begin{eqnarray}
P_s^{(N,E)}\bigotimes \{ L \}  &\ \ & P=(-1)^{L+1},\ \ \ C=\ \ (-1)^L\nonumber\\
V_\mu^{(N,E)}\bigotimes \{ L \}  &\ \ & P=(-1)^{L+1},\ \ \ C=\ \ (-1)^{L+1}\nonumber\\
S^{(N,E)}\bigotimes \{ L \}  &\ \ & P=(-1)^{L},\ \ \ C=\pm (-1)^L\nonumber\\
A_\mu^{(N,E)}\bigotimes \{ L \}  &\ \ & P=(-1)^{L},\ \ \ C=\pm (-1)^L  .
\end{eqnarray}

\subsection{Expected spectroscopy of mesons}
In our fundamental equations Eqs.(\ref{eq1}) to (\ref{eq2c})
it was supposed that the squared-mass spectra ${\cal M}^2$ is, as a first step  
in the pure-confining force limit, Dirac-spinor independent and also quark-flavor
independent.\footnote{
Of course we consider, separately, the light-quark system, light/heavy-quark system
and heavy-quark system. 
}
Actually we must take into account\cite{rfmass} the various effects due to one-gluon-exchange
potential\footnote{
We must also take the other non-perturbative QCD effects like quark-condensation and 
instantons.
} and also the effects due to quark mass difference.\\
$\underline{light-quark\ (q\bar q)\ meson\ system}$:\ \ \  In the pure-confining force limit
all the ground state mesons expected in \S 3.1 , $P_s^{(N,E)}, V_\mu^{(N,E)}, S^{(N,E)}$
and $A_\mu^{(N,E)}$ are degenerate and have the same mass, $M_0=m_1+m_2$.
Actually the mass of $P_s^{(N)}$, to be assigned $\pi$-nonet, should be exceptionally
low because of its nature\footnote{
The $q\bar q$-condensation corresponds to the non-zero vacuum expectation value of
$S^{(N)}$, $\langle S^{(N)} \rangle_0$. Thus $P_s^{(N)}$ other than $P_s^{(E)}$ is
a Nambu-Goldstone boson.
}   
as a Nambu-Goldstone boson. The masses of the $V_\mu^{(N)}$-nonet, which is assigned
to be $\rho$-meson nonet, are almost equal to
the corresponding masses $M_0=m_1+m_2$ in the pure-confining force limit. 
The masses of all the other ground-state mesons
are expected to be almost equal to those of the corresponding normal vector mesons, 
and to be lower than those of the corresponding first-excited states.

For the first-excited states, the chiral symmetry is expected to be still effective,
as is seen from Table II given in the last subsection \S 5.2 , so we expect the
existence of a series of the first excited $P$-wave states of the ground state multiplets.
They are expected to have the masses, which are almost equal to the first excited states
of normal vector mesons, and which are lower than the second excited states of those. 
Among the multiplets newly predicted in the present scheme, to be
called ``chiralons," the especially interesting mesons are the ones with 
$J^{PC}$=$0^{+-}(S^{(E)}(S-wave)),\ 1^{-+}(S^{(E)}(P-wave) )$ and 
$1^{-+}(A_\mu^{(E)}(P-wave))$, which are ``exotic particles" out of 
the conventional 
non-relativistic $q\bar q$-mesons. 
Their masses, by the above mentioned estimate,
is expected to be, respectively, 
\begin{eqnarray} 
m(0^{+-}) & \stackrel{<}{\scriptstyle \sim}& 1.3\ {\rm GeV},\nonumber\\
1.3\ {\rm GeV} & \simeq & m(1^{-+},S^{(E)}) 
                 \simeq  m(1^{-+},A_\mu^{(E)}) 
    \stackrel{<}{\scriptstyle \sim}  1.7\ {\rm GeV}
\label{eq55}
\end{eqnarray}\\
$\underline{heavy-light\ quark\ (Q\bar q\ and\ q\bar Q)\ meson\ system}$:\ \ \  As is seen 
from Table I we are able to expect the existence of new multiplets (at least the ground
states of scalar and axial-vector triplet).\\
\ \ \ \ \\ 
$\underline{heavy\ quark\ (Q\bar Q)\ meson\ system}$:\ \ \  No new multiplets are expected
to exist. \\


\begin{table}
\caption{Quantum numbers of Ground and Excited states 
of light through heavy quark-antiquark mesons
and their experimental candidates.
For $Q\bar q,\ q\bar Q$-system, angular momentum of 
the ``light degrees of freedom" $j_q$,
which are obtained by ${\mib j}_q={\mib L}+{\mib S}_q$, is expected to be 
good quantum number, and the two $P$-wave meson states with 
$J=1$, $j_q=1/2$ and $3/2$,
$M(P_1^{ j_q=1/2,\ 3/2})$ are the linear combinations of the conventional 
$LS$-states, $M(^{1,3}P_1)$, as
$M(P_1^{j_q=1/2})=\sqrt{2/3}M(^3P_1)-\sqrt{1/3}M(^1P_1)$, 
and 
$M(P_1^{j_q=3/2})=-\sqrt{1/3}M(^3P_1)-\sqrt{2/3}M(^1P_1)$.
} 
\begin{center}
\begin{tabular}{|l|cc|l|}
\hline\hline
$q\bar q$ & & $J^{PC}$ & Experimental candidates\\
\hline
$P_s^{(N)}\bigotimes \{ L \}$ & $^1S_0$ & $0^{-+}$ & $\pi , K, \eta ,\eta '$ \\
$V_\mu^{(N)}\bigotimes \{ L \}$ & $^3S_1$ & $1^{--}$ & $\rho , K^*, \omega ,\phi$ \\
      & $^1P_1$ & $1^{+-}$ & $h_1(1170), b_1(1230)$ \\
      & $^3P_0$ & $0^{++}$ & $f_0(1370),K_0^*(1430),a_0(1450),(f_0(1500))$ \\
      & $^3P_1$ & $1^{++}$ & $a_1(1230),f_1(1285),K_1(1270)$ \\
      & $^3P_2$ & $2^{++}$ & $a_2(1320),f_2(1275),K_2^*(1430),f_2'(1525)$ \\
\hline
$P_s^{(E)}\bigotimes \{ L \}$ & $^1S_0$ & $0^{-+}$
   & (one out of $\eta (1275),\eta (1420)$ and $\eta (1460))$ \\
$V_\mu^{(E)}\bigotimes \{ L \}$ & $^3S_1$ & $1^{--}$ &  \\
      & $^1P_1$ & $1^{+-}$ &  \\
      & $^3P_0$ & $0^{++}$ &  \\
      & $^3P_1$ & $1^{++}$ &  \\
      & $^3P_2$ & $2^{++}$ &  \\
\hline
$S^{(N)}\bigotimes \{ L \}$ & $^1S_0$ & $0^{++}$ & $\sigma ,
\kappa ,a_0(980)$=$\delta , f_0(980)$=$\sigma '$ \\
      & $^1P_1$ & $1^{--}$ &   \\
$A_\mu^{(N)}\bigotimes \{ L \}$ & $^3S_1$ & $1^{++}$ & $a_1(900)$ \\
      & $^3P_0$ & $0^{--}$ &  \\
      & $^3P_1$ & $1^{--}$ &  \\
      & $^3P_2$ & $2^{--}$ &  \\
\hline
$S^{(E)}\bigotimes \{ L \}$ & $^1S_0$ & $\underline{0^{+-}}$ &  \\
      & $^1P_1$ & $\underline{1^{-+}}$ &  $\pi_1(1400)$  \\
$A_\mu^{(E)}\bigotimes \{ L \}$ & $^3S_1$ & $1^{+-}$ &   \\
      & $^3P_0$ & $0^{-+}$ &  \\
      & $^3P_1$ & $\underline{1^{-+}}$ & $\pi_1(1600)$ \\
      & $^3P_2$ & $2^{-+}$ &  \\
\hline\hline
$Q\bar q,\ q\bar Q$ & & $J^P$ &  \\
\hline
$P_s\bigotimes\{ L \}$ & $^1S_0$ & $0^-$ & $D,\ B,\ D_s,\ B_s$ \\
$V_\mu\bigotimes\{ L \}$ & $^3S_1$ & $1^-$ & $D^*,\ B^*,\ D_s^*,\ B_s^*$ \\
  & $P_1^{j_q=1/2}$ & $1^+$ & $D_1^*$ \\
  & $P_1^{j_q=3/2}$ & $1^+$ & $D_1$ \\
  & $^3P_0$      & $0^+$ & $D_0^*$ \\
  & $^3P_2$      & $2^+$ & $D_2^*$ \\
\hline
$S\bigotimes\{ L \}$     & $^1S_0$ & $0^+$ & $B_0^c(5520)$ \\
$A_\mu\bigotimes\{ L \}$ & $^3S_1$ & $1^+$ &  \\
  & $P_1^{j_q=1/2}$ & $1^-$ &  \\
  & $P_1^{j_q=3/2}$ & $1^-$ &  \\
  & $^3P_0$      & $0^-$ &  \\
  & $^3P_2$      & $2^-$ &  \\
\hline\hline
$Q\bar Q$ & & $J^P$  &  \\
\hline
$P_s\bigotimes\{ L \}$ & $^1S_0$ & $0^-$ & $\eta_c,\ B_c$ \\
$V_\mu\bigotimes\{ L \}$ & $^3S_1$ & $1^-$ & $J/\psi,\ \Upsilon ,\  B_c^*$ \\
  & $^1P_1$ & $1^+$ & $h_c$ \\
  & $^3P_0$ & $0^+$ & $\chi_{c0},\  \chi_{b0}$ \\
  & $^3P_1$ & $1^+$ & $\chi_{c1},\  \chi_{b1}$ \\
  & $^3P_2$ & $2^+$ & $\chi_{c2},\  \chi_{b2}$ \\
\hline\hline
\end{tabular}
\end{center}
\end{table}

\section{Experimental Evidences and Concluding Remarks}
In this paper we have presented a kinematical framework for describing 
covariantly the ground states as well as excited states of light-through-heavy
quark mesons. For light-quark mesons our scheme gives a theoretical basis
to classify the composite meson systems unifying the two contrasting viewpoints 
based on non-relativistic quark model with $LS$ symmetry and on NJL model
with chiral symmetry. The essential physical assumption is to set up the 
Klein-Gordon type of Yukawa equation on the bi-local meson wave function with
the squared-mass operator, which is, in the pure-confining force limit, independent
of Dirac-spinor suffix, and accordingly is chiral symmetric.
As a result is pointed out a possibility of existence of 
rather an abundant new nonets, chiralons, 
with masses lower than about 2 GeV; several new ground 
state meson nonets and some new excited meson nonets.

For heavy/light quark meson systems 
we have similarly pointed out a possibility
of existence of new multiplets(triplets), chiralons.

In Table III we have summarized the expected level structure of general ground and 
the first-excited quark-antiquark mesons. 
Presently we can give a few experimental candidates for the predicted 
members of new multiplets: One of the most important and interesting ones is
the scalar $\sigma$ nonet; the members\footnote{
This assignment of $\sigma$ nonet was first proposed in Ref. \citen{rfsca}
and afterwards in Refs. \citen{rfRupp,rftorn}, 
and by one of  the present authors in Ref. \citen{rfmy}.
Rather strong experimental evidences for $\sigma$(600) have been given recently
by many authors; from the $\pi\pi$ scattering process 
Refs. \citen{rfpipi1,rfpipi2,rfpipi3,rfpipi4,rfpipi5} .
and from the $\pi\pi$ production processes Ref. \citen{rfprod1,rfprod2,rfprod3,rfprod4}.   
Possible evidences for $\kappa$(900) was pointed out by reanalyzing the $K\pi$ 
scattering phase shift in Refs. \citen{rfRupp}, \citen{rfkappa} and \citen{rfsche}.
}
are $\sigma (600)$, $\kappa$(900), $a_0(980)$ and $f_0(980)$, which constitute,
with the members of $\pi$-nonet, a linear representation of the chiral $SU(3)$ symmetry.
It is notable that the $\sigma$ nonet is the relativistic $S$-wave states, which should 
be discriminated from the non-relativistic $^3P_0$ states.

Another example suggesting possible validity of 
the present scheme is existence
of the three pseudoscalars with mass between 1 GeV$\sim$1.5GeV, 
$\eta$(1295),\cite{rfeta1,rfeta2}\\
$\eta$(1420)\cite{rfeta3,rfeta1,rfeta4} and $\eta$(1460).\cite{rfeta4} 
The two out of them may be the members of the radially
excited $\pi$-nonet, while the one extra may belong to the ground states of the extra
pseudoscalar nonet newly predicted.

Also we have the other candidates for chiralons: It was a problem for experimental groups
for long time whether a possible resonance observed in the $\eta\pi$ system, with an exotic
quantum number $J^{PC}=1^{-+}$ and with a mass around 1.5 GeV, really exists or does not.
Recently it seems that the existence of two such particles\cite{rf6} $\pi_1(1400)$  and
$\pi_1(1600)$ have been accepted\cite{rfchung} widely.
These two particles have the mass in the region estimated in Eq.(\ref{eq55})
to be assigned as the respective excited $P$-wave states of $S^{(E)}$ and $A_\mu^{(E)}$.

We have the other longstanding problem in hadron spectroscopy: the mass and width of
$a_1(1260)$ seem to be variant\cite{rfR1} depending\cite{rfR2}
 on the production process and/or decay channel.

In connection to this problem we have made recently a preliminary 
analysis\cite{rfwaka} of
the data\cite{rfkoba} obtained by GAMS group WA102 experiment on process
$\pi^- p\to 3\pi^0 n$. As a result we have obtained an evidence of existence of two
$a_1(J^{PC}=1^{++})$ particles: $a_1^c(m=0.9{\rm GeV},\Gamma =200{\rm MeV})$,
and $a_1^N(m=1.2{\rm GeV},\Gamma =440{\rm MeV})$.
The former may be assigned to be $A_\mu^{(N)} (^3S_1)$, 
while the latter be conventional
$a_1$ particle (to be $V_\mu^{(N)}(^3P_1)$ in our classification scheme).

All the above candidates for chiralons are concerned 
with the light-quark mesons.
Here, we should like to refer to a preliminary result concerning the 
heavy/light quark meson systems that a scalar chiralons
$B_0^c$ with $M$=5.52 GeV and $\Gamma$=44 MeV may be observed\cite{rfR3}
in the $B\pi$ channel produced\cite{rfR4} through the $Z$-boson decay.

We have also given 
the above mentioned experimental candidates for chiralons
in Table III, where 
we have listed, for reference, also our assignment to the conventional
$S$-wave and $P$-wave states.

In concluding we should like to remark that 
in this paper we have dared to present a very ``brave" attempt for unified classification
scheme of mesons, 
which predicts the possible existence of a lot of new meson multiplets.
Further serious investigations and search for them 
will be required to test validity of the present scheme.

\acknowledgements

The authors would like to express their sincere gratitude to 
prof. K. Takamatsu, T. Tsuru and T. Sawada for encouragements 
and useful informations. 
They would like to thank prof. K. Yamada for useful comments.
They are also grateful to Dr. T. Ishida for encouragements and comments. 
 
\appendix
\section{Spin wave functions of composite meson systems\\
 and fundamental crossing rules for constituent quarks}
In the free local field its annihilation (positive frequency)
and creation (negative frequency)
parts are related with each other by the
conventional crossing rule. 
In this appendix we shall derive a similar crossing rule for 
our covariant spin WF of composite meson systems by applying the 
fundamental crossing rule\cite{rf31} of our 
extended Dirac spinors for constituent quarks. 
In the following 
we first make the annihilation part of the composite meson WF 
by decomposition of total spin of the constituent quarks and antiquarks. 
Next by using the fundamental crossing rules for the constituent spinors  
we construct the creation part of the composite meson spinor WF.

We use the following conventional  ``free" Dirac $u$ and $v$ spinors: 
\begin{eqnarray}
u({\mib p},h) &=& \left(  \begin{array}{l} \sqrt{E+m}\chi^{(h)} \\
                             \sqrt{E-m}{\mib n}\cdot{\mib \sigma}\chi^{(h)}
                          \end{array} \right) ,\ \ \ 
v({\mib p},h) = \left(  \begin{array}{l}  \sqrt{E-m}
                            {\mib n}\cdot{\mib \sigma}\chi^{(h)\prime} \\
                             \sqrt{E+m}\chi^{(h)\prime}  
                          \end{array} \right) .
\end{eqnarray}
In this Appendix A 
we choose the $z$-axis parallel to the momentum of composite mesons as 
${\mib p}=p{\mib n}=p\hat{\mib z}$, 
and accordingly
\begin{eqnarray}
\chi^{(+)} &=& \left( \begin{array}{c}1\\ 0\end{array}\right) ,
\chi^{(-)} = \left( \begin{array}{c}0\\ 1\end{array}\right) ,\ \ 
\chi^{(h)\prime}\equiv -i\sigma_2\chi^{(h)*},
\begin{array}{c} \chi^{(+)\prime} \\  \chi^{(-)\prime} \end{array}
\begin{array}{c} = \\  = \end{array}
\begin{array}{r} \chi^{(-)} \\  -\chi^{(+)} \end{array} .
\end{eqnarray}

\subsection{quark and antiquark spinor inside of mesons}
Here we give the annihilation part of the constituent quark and antiquark 
spinor WF. 
As was explained in the text 
the constituent quark inside of mesons 
has the freedom of positive and negative energy, 
as well as that of up and down spin, and totally four degrees of freedom.

The positive energy quark spinor $u_+$ 
with the meson momentum ${\mib p}$, 
which is related with the ``free" Dirac $u$
spinor, is defined by
\begin{eqnarray}
u_+({\mib p},\pm ) &=& \left(  \begin{array}{l} \sqrt{E+m}\chi^{(\pm )} \\
                   \sqrt{E-m}{\mib n}\cdot{\mib \sigma}\chi^{(\pm )}
              \end{array} \right)  = u({\mib p},\pm ) .
\end{eqnarray}
The negative energy quark spinor $u_-$ 
with the momentum, $-{\mib p}$, anti-parallel to the meson momentum ${\mib p}$,
is defined 
by using the same spin operator ${\mib n}\cdot{\mib \sigma}$ as for $u_+$. 
They are related with free ``Dirac" $v$ spinors.
\begin{eqnarray}
u_-(-{\mib p},\pm ) &=& \left(  \begin{array}{l}  \sqrt{E-m}
                            {\mib n}\cdot{\mib \sigma}\chi^{(\pm )} \\
                             \sqrt{E+m}\chi^{(\pm )}  
                          \end{array} \right) 
=\mp v({\mib p},\mp ).
\end{eqnarray}

The positive energy antiquark spinor $\bar v_+$ with momentum ${\mib p}$,
which is related with the ``free" Dirac $\bar v$
spinor, is defined by
\begin{eqnarray}
\bar v_+({\mib p},\pm ) &=&  (\sqrt{E-m}\chi^{(\pm )\prime\dagger}{\mib n}\cdot{\mib \sigma},
  -\sqrt{E+m}\chi^{(\pm )\prime\dagger}) = \bar v({\mib p},\pm ) .
\end{eqnarray}
The negative energy antiquark spinor $\bar v_-$ with momentum $-{\mib p}$, 
is defined, similarly by using the same spin operator 
$-{\mib n}\cdot{\mib \sigma}$, 
as in the case of $\bar v_+$. 
They are related with free ``Dirac" $\bar u$ spinors.
\begin{eqnarray}
\bar v_-(-{\mib p},\pm ) &=&  
(\sqrt{E+m}\chi^{(\pm )\prime\dagger},
-\sqrt{E-m}\chi^{(\pm )\prime\dagger}{\mib n}\cdot{\mib \sigma}) 
= \pm \bar u({\mib p},\mp ) .
\end{eqnarray}

\subsection{Fundamental crossing rules for constituent spinor inside of mesons}

Under the crossing operation for mesons the 
annhilation (positive frequency) part of meson WF 
is transformed into the creation 
(negative frequency) part of that.
The corresponding fundamental crossing rule is set up as follows:
For example,  
 the positive energy constituent quark spinor in the annihilation part of meson  WF 
is transformed into the positive energy antiquark spinor in the 
the creation part of meson WF. 
\begin{eqnarray}
u_+ ({\mib p},s) &\longrightarrow&  v_+ ({\mib p},s)\ \ \ \ 
{\rm that}\ {\rm is}\ \ u({\mib p},\pm )\longrightarrow v({\mib p},\pm ) .
\label{eqcr1}
\end{eqnarray}
The other kinds of constituent quark spinors are transformed similarly as 
\begin{eqnarray}
u_-(-{\mib p},s) &\longrightarrow&  v_- (-{\mib p},s)\ \ \ \ 
{\rm that}\ {\rm is}\ \ \mp v({\mib p},\mp )\longrightarrow \pm u({\mib p},\mp ) ,
\label{eqcr2}
\end{eqnarray}
%
\begin{eqnarray}
\bar v_+ ({\mib p},s) &\longrightarrow&  \bar u_+ ({\mib p},s)\ \ \ \ 
{\rm that}\ {\rm is}\ \ \bar v({\mib p},\pm )\longrightarrow 
\bar u({\mib p},\pm ) ,
\label{eqcr3}
\end{eqnarray}
\begin{eqnarray}
\bar v_-(-{\mib p},s) &\longrightarrow&  \bar u_- (-{\mib p},s)\ \ \ \ 
{\rm that}\ {\rm is}\ \ \pm \bar u({\mib p},\mp )\longrightarrow \mp 
\bar v({\mib p},\mp ) .
\label{eqcr4}
\end{eqnarray}

\subsection{Annihilation part of the composite meson spinor WF}
As was explained in the text, depending upon the energy sign of 
the constituent spinor WF, 
there are four different types of
(the annihilation part of) bi-spinor WF, 
$U^{(+)}$, $C^{(+)}$, $D^{(+)}$ and $V^{(+)}$.
Each type of the WF is decomposed into the irreducible components, which are
constructed by using the usual spin composition.

The $U^{(+)} \sim u_+\bar v_+ (\sim u\bar v)$ is decomposed into the pseudoscalar and vector 
components, which are constructed from the constituent spinors as
\begin{eqnarray}
U^{(+)}_{P_s} &=& \frac{1}{i} \frac{1}{2m} \frac{1}{\sqrt 2} 
 (u_+({\mib p},+)\bar v_+({\mib p},-)-u_+({\mib p},-)\bar v_+({\mib p},+))  \\
 &=& \frac{1}{i} \frac{1}{2m} \frac{1}{\sqrt 2} 
 (u({\mib p},+)\bar v({\mib p},-)-u({\mib p},-)\bar v({\mib p},+)) 
 = \frac{1}{2\sqrt 2} i\gamma_5 (1+iv\cdot\gamma ) \nonumber\\
U^{(+)}_{V_\mu^{(\pm ,0)} } &=& \frac{1}{2m} \left[ \begin{array}{l}
u_+({\mib p},+)\bar v_+({\mib p},+) \\  u_+({\mib p},-)\bar v_+({\mib p},-) \\
\frac{1}{\sqrt 2}  (u_+({\mib p},+)\bar v_+({\mib p},-)+u_+({\mib p},-)\bar v_+({\mib p},+)) 
\end{array} \right] \nonumber\\
&=&  \frac{1}{2m} \left[ \begin{array}{l}
u({\mib p},+)\bar v({\mib p},+) \\  u({\mib p},-)\bar v({\mib p},-) \\
\frac{1}{\sqrt 2}  (u({\mib p},+)\bar v({\mib p},-)+u({\mib p},-)\bar v({\mib p},+)) 
\end{array} \right]     \\
&=& \frac{1}{4\sqrt 2} (1-iv\cdot\gamma )i\epsilon^{(\pm ,0)}\cdot\gamma (1+iv\cdot\gamma )
=\frac{1}{2\sqrt 2} i\epsilon^{(\pm ,0)}\cdot\gamma (1+iv\cdot\gamma ) .\nonumber
\end{eqnarray}

The $C^{(+)} \sim u_+\bar v_- (\sim u\bar u)$ is decomposed into the scalar and axialvector 
components, which are constructed from the constituent spinors as
\begin{eqnarray}
C^{(+)}_{S} &=& - \frac{1}{2m} \frac{1}{\sqrt 2} 
 (u_+({\mib p},+)\bar v_-(-{\mib p},-)-u_+({\mib p},-)\bar v_-(-{\mib p},+))  \\
 &=& - \frac{1}{2m} \frac{1}{\sqrt 2} 
 (-u({\mib p},+)\bar u({\mib p},+)-u({\mib p},-)\bar u({\mib p},-)) 
 = \frac{1}{2\sqrt 2} (1-iv\cdot\gamma )   \nonumber\\
C^{(+)}_{A_\mu^{(\pm ,0)} } &=& -\frac{1}{2m} \left[ \begin{array}{l}
u_+({\mib p},+)\bar v_-(-{\mib p},+) \\  u_+({\mib p},-)\bar v_-(-{\mib p},-) \\
\frac{1}{\sqrt 2}  (u_+({\mib p},+)\bar v_-(-{\mib p},-)+u_+({\mib p},-)\bar v_-(-{\mib p},+)) 
\end{array} \right] \nonumber\\
&=&  -\frac{1}{2m} \left[ \begin{array}{l}
u({\mib p},+)\bar u({\mib p},-) \\  -u({\mib p},-)\bar u({\mib p},+) \\
\frac{1}{\sqrt 2}  (-u({\mib p},+)\bar u({\mib p},+)+u({\mib p},-)\bar u({\mib p},-)) 
\end{array} \right]     \\
&=& \frac{1}{4\sqrt 2} (1-iv\cdot\gamma )i\gamma_5\epsilon^{(\pm ,0)}\cdot\gamma 
(1-iv\cdot\gamma )
=\frac{1}{2\sqrt 2} i\gamma_5\epsilon^{(\pm ,0)}\cdot\gamma (1-iv\cdot\gamma ) .\nonumber
\end{eqnarray}

The $D^{(+)} \sim u_-\bar v_+ (\sim v\bar v)$ is decomposed into the scalar and axialvector 
components, which are constructed from the constituent spinors as
\begin{eqnarray}
D^{(+)}_{S} &=& \frac{1}{2m} \frac{1}{\sqrt 2} 
 (u_-(-{\mib p},+)\bar v_+({\mib p},-)-u_-(-{\mib p},-)\bar v_+({\mib p},+))  \\
 &=& \frac{1}{2m} \frac{1}{\sqrt 2} 
 (-v({\mib p},-)\bar v({\mib p},-)-v({\mib p},+)\bar v({\mib p},+)) 
 = \frac{1}{2\sqrt 2} (1+iv\cdot\gamma ) \nonumber\\
D^{(+)}_{A_\mu^{(\pm ,0)} } &=& -\frac{1}{2m} \left[ \begin{array}{l}
u_-(-{\mib p},+)\bar v_+({\mib p},+) \\  u_-(-{\mib p},-)\bar v_+({\mib p},-) \\
\frac{1}{\sqrt 2}  (u_-(-{\mib p},+)\bar v_+({\mib p},-)+u_-(-{\mib p},-)\bar v_+({\mib p},+)) 
\end{array} \right] \nonumber\\
&=&  -\frac{1}{2m} \left[ \begin{array}{l}
-v({\mib p},-)\bar v({\mib p},+) \\  v({\mib p},+)\bar v({\mib p},-) \\
\frac{1}{\sqrt 2}  (-v({\mib p},-)\bar v({\mib p},-)+v({\mib p},+)\bar v({\mib p},+)) 
\end{array} \right]     \\
&=& \frac{1}{4\sqrt 2} (1+iv\cdot\gamma )i\gamma_5\epsilon^{(\pm ,0)}\cdot\gamma 
(1+iv\cdot\gamma )
=\frac{1}{2\sqrt 2} i\gamma_5\epsilon^{(\pm ,0)}\cdot\gamma (1+iv\cdot\gamma ) .\nonumber
\end{eqnarray}

The $V^{(+)} \sim u_-\bar v_- (\sim v\bar u)$ is decomposed into the pseudoscalar and vector 
components, which are constructed from the constituent spinors as
\begin{eqnarray}
V^{(+)}_{P_s} &=& i \frac{1}{2m} \frac{1}{\sqrt 2} 
 (u_-(-{\mib p},+)\bar v_-(-{\mib p},-)-u_-(-{\mib p},-)\bar v_-(-{\mib p},+))  \\
 &=& i \frac{1}{2m} \frac{1}{\sqrt 2} 
 ( v({\mib p},-)\bar u({\mib p},+)-v({\mib p},+)\bar u({\mib p},-)) 
 = \frac{1}{2\sqrt 2}i\gamma_5 (1-iv\cdot\gamma ) \nonumber\\
V^{(+)}_{V_\mu^{(\pm ,0)} } &=& \frac{1}{2m} \left[ \begin{array}{l}
u_-(-{\mib p},+)\bar v_-(-{\mib p},+) \\  u_-(-{\mib p},-)\bar v_-(-{\mib p},-) \\
\frac{1}{\sqrt 2}  (u_-(-{\mib p},+)\bar v_-(-{\mib p},-)+u_-(-{\mib p},-)\bar v_-(-{\mib p},+)) 
\end{array} \right] \nonumber\\
&=&  \frac{1}{2m} \left[ \begin{array}{l}
-v({\mib p},-)\bar u({\mib p},-) \\  -v({\mib p},+)\bar u({\mib p},+) \\
\frac{1}{\sqrt 2}  (v({\mib p},-)\bar u({\mib p},+)+v({\mib p},+)\bar u({\mib p},-)) 
\end{array} \right]     \\
&=& \frac{1}{4\sqrt 2} (1+iv\cdot\gamma )i\epsilon^{(\pm ,0)}\cdot\gamma 
(1-iv\cdot\gamma )
=\frac{1}{2\sqrt 2} i\epsilon^{(\pm ,0)}\cdot\gamma (1-iv\cdot\gamma ) .\nonumber
\end{eqnarray}

\subsection{Creation part of the composite meson spinor WF}
Creation part of composite meson spinor WF is obtained 
from the annihilation part, by applying the fundamental  
crossing rule Eqs.(\ref{eqcr1}), (\ref{eqcr2}),
(\ref{eqcr3}) and (\ref{eqcr4}), as follows:\\
For $U^{(-)} \sim v_+\bar u_+ (\sim v\bar u)$ 
\begin{eqnarray}
U^{(-)}_{P_s} &=& \frac{1}{i} \frac{1}{2m} \frac{1}{\sqrt 2} 
 (v_+({\mib p},+)\bar u_+({\mib p},-)-v_+({\mib p},-)\bar u_+({\mib p},+))  \\
 &=& \frac{1}{i} \frac{1}{2m} \frac{1}{\sqrt 2} 
 (v({\mib p},+)\bar u({\mib p},-)-v({\mib p},-)\bar u({\mib p},+)) 
 = \frac{1}{2\sqrt 2} i\gamma_5 (1-iv\cdot\gamma ) \nonumber\\
U^{(-)}_{V_\mu^{(\pm ,0)} } &=& \frac{1}{2m} \left[ \begin{array}{l}
v_+({\mib p},+)\bar u_+({\mib p},+) \\  v_+({\mib p},-)\bar u_+({\mib p},-) \\
\frac{1}{\sqrt 2}  (v_+({\mib p},+)\bar u_+({\mib p},-)+v_+({\mib p},-)\bar u_+({\mib p},+)) 
\end{array} \right] \nonumber\\
&=&  \frac{1}{2m} \left[ \begin{array}{l}
v({\mib p},+)\bar u({\mib p},+) \\  v({\mib p},-)\bar u({\mib p},-) \\
\frac{1}{\sqrt 2}  (v({\mib p},+)\bar u({\mib p},-)+v({\mib p},-)\bar u({\mib p},+)) 
\end{array} \right]     \\
&=& \frac{1}{4\sqrt 2} (1+iv\cdot\gamma )i\tilde\epsilon^{(\pm ,0)}\cdot\gamma (1-iv\cdot\gamma )
=\frac{1}{2\sqrt 2} i\tilde\epsilon^{(\pm ,0)}\cdot\gamma (1-iv\cdot\gamma ) .\nonumber
\end{eqnarray}
For $C^{(-)} \sim v_+\bar u_- (\sim v\bar v)$ 
\begin{eqnarray}
C^{(-)}_{S} &=& - \frac{1}{2m} \frac{1}{\sqrt 2} 
 (v_+({\mib p},+)\bar u_-(-{\mib p},-)-v_+({\mib p},-)\bar u_-(-{\mib p},+))  \\
 &=& - \frac{1}{2m} \frac{1}{\sqrt 2} 
 (v({\mib p},+)\bar v({\mib p},+)+v({\mib p},-)\bar v({\mib p},-)) 
 = \frac{1}{2\sqrt 2} (1+iv\cdot\gamma )   \nonumber\\
C^{(-)}_{A_\mu^{(\pm ,0)} } &=& -\frac{1}{2m} \left[ \begin{array}{l}
v_+({\mib p},+)\bar u_-(-{\mib p},+) \\  v_+({\mib p},-)\bar u_-(-{\mib p},-) \\
\frac{1}{\sqrt 2}  (v_+({\mib p},+)\bar u_-(-{\mib p},-)+v_+({\mib p},-)\bar u_-(-{\mib p},+)) 
\end{array} \right] \nonumber\\
&=&  -\frac{1}{2m} \left[ \begin{array}{l}
-v({\mib p},+)\bar v({\mib p},-) \\  v({\mib p},-)\bar v({\mib p},+) \\
\frac{1}{\sqrt 2}  (v({\mib p},+)\bar v({\mib p},+)-v({\mib p},-)\bar v({\mib p},-)) 
\end{array} \right]     \\
&=& \frac{1}{4\sqrt 2} (1+iv\cdot\gamma )i\gamma_5\tilde\epsilon^{(\pm ,0)}\cdot\gamma 
(1+iv\cdot\gamma )
=\frac{1}{2\sqrt 2} i\gamma_5\tilde\epsilon^{(\pm ,0)}\cdot\gamma (1+iv\cdot\gamma ) .\nonumber
\end{eqnarray}
For $D^{(-)} \sim v_-\bar u_+ (\sim u\bar u)$ 
\begin{eqnarray}
D^{(-)}_{S} &=& \frac{1}{2m} \frac{1}{\sqrt 2} 
 (v_-(-{\mib p},+)\bar u_+({\mib p},-)-v_-(-{\mib p},-)\bar u_+({\mib p},+))  \\
 &=& \frac{1}{2m} \frac{1}{\sqrt 2} 
 (u({\mib p},-)\bar u({\mib p},-)+u({\mib p},+)\bar u({\mib p},+)) 
 = \frac{1}{2\sqrt 2} (1-iv\cdot\gamma ) \nonumber\\
D^{(-)}_{A_\mu^{(\pm ,0)} } &=& -\frac{1}{2m} \left[ \begin{array}{l}
v_-(-{\mib p},+)\bar u_+({\mib p},+) \\  v_-(-{\mib p},-)\bar u_+({\mib p},-) \\
\frac{1}{\sqrt 2}  (v_-(-{\mib p},+)\bar u_+({\mib p},-)+v_-(-{\mib p},-)\bar u_+({\mib p},+)) 
\end{array} \right] \nonumber\\
&=&  -\frac{1}{2m} \left[ \begin{array}{l}
u({\mib p},-)\bar u({\mib p},+) \\  -u({\mib p},+)\bar u({\mib p},-) \\
\frac{1}{\sqrt 2}  (u({\mib p},-)\bar u({\mib p},-)-u({\mib p},+)\bar u({\mib p},+)) 
\end{array} \right]     \\
&=& \frac{1}{4\sqrt 2} (1-iv\cdot\gamma )i\gamma_5\tilde\epsilon^{(\pm ,0)}\cdot\gamma 
(1-iv\cdot\gamma )
=\frac{1}{2\sqrt 2} i\gamma_5\tilde\epsilon^{(\pm ,0)}\cdot\gamma (1-iv\cdot\gamma ) .\nonumber
\end{eqnarray}
For $V^{(-)} \sim v_-\bar u_- (\sim u\bar v)$
\begin{eqnarray}
V^{(-)}_{P_s} &=& i \frac{1}{2m} \frac{1}{\sqrt 2} 
 (v_-(-{\mib p},+)\bar u_-(-{\mib p},-)-v_-(-{\mib p},-)\bar u_-(-{\mib p},+))  \\
 &=& i \frac{1}{2m} \frac{1}{\sqrt 2} 
 ( u({\mib p},-)\bar v({\mib p},+)-u({\mib p},+)\bar v({\mib p},-)) 
 = \frac{1}{2\sqrt 2}i\gamma_5 (1+iv\cdot\gamma ) \nonumber\\
V^{(-)}_{V_\mu^{(\pm ,0)} } &=& \frac{1}{2m} \left[ \begin{array}{l}
v_-(-{\mib p},+)\bar u_-(-{\mib p},+) \\  v_-(-{\mib p},-)\bar u_-(-{\mib p},-) \\
\frac{1}{\sqrt 2}  (v_-(-{\mib p},+)\bar u_-(-{\mib p},-)+v_-(-{\mib p},-)\bar u_-(-{\mib p},+)) 
\end{array} \right] \nonumber\\
&=&  \frac{1}{2m} \left[ \begin{array}{l}
-u({\mib p},-)\bar v({\mib p},-) \\  -u({\mib p},+)\bar v({\mib p},+) \\
\frac{1}{\sqrt 2}  (u({\mib p},-)\bar v({\mib p},+)+u({\mib p},+)\bar v({\mib p},-)) 
\end{array} \right]     \\
&=& \frac{1}{4\sqrt 2} (1-iv\cdot\gamma )i\tilde\epsilon^{(\pm ,0)}\cdot\gamma 
(1+iv\cdot\gamma )
=\frac{1}{2\sqrt 2} i\tilde\epsilon^{(\pm ,0)}\cdot\gamma (1+iv\cdot\gamma ) .\nonumber
\end{eqnarray}

\subsection{Representation by local meson field}
We can simply combine the annihilation and creation parts of 
the composite spinor WF
into the local meson field(, neglecting the freedom of internal space-time $x$).\\
For non-relativistic type spinors
\begin{eqnarray}
\frac{1}{2\sqrt 2} &i& 
\gamma_5 (1+\gamma_\nu\frac{\partial_\nu}{\sqrt{\partial^2}}) P_s^{(NR)}(X), \ \ \ 
\frac{1}{2\sqrt 2} i
\gamma_\mu (1+\gamma_\nu\frac{\partial_\nu}{\sqrt{\partial^2}}) V_\mu^{(NR)}(X) ,
\end{eqnarray}    
where $P_s^{(NR)}(X)$ and $V_\mu^{(NR)}(X)$ represents the local pseudoscalar
and vector meson field operators of non-relativistic-type, respectively.\\
For relativistic $\bar q$-type spinors
\begin{eqnarray}
\frac{1}{2\sqrt 2} &(& 
1-\gamma_\nu\frac{\partial_\nu}{\sqrt{\partial^2}}) S^{\bar q}(X), \ \ \ 
\frac{1}{2\sqrt 2} i
\gamma_5\gamma_\mu (1-\gamma_\nu\frac{\partial_\nu}{\sqrt{\partial^2}}) 
A_\mu^{\bar q}(X) ,
\end{eqnarray}    
where $S^{\bar q}(X)$ and $A_\mu^{\bar q}(X)$ represents the local pseudoscalar
and vector meson field operators of $\bar q$-type, respectively.\\
For relativistic $q$-type spinors
\begin{eqnarray}
\frac{1}{2\sqrt 2} &(& 
1+\gamma_\nu\frac{\partial_\nu}{\sqrt{\partial^2}}) S^q(X), \ \ \ 
\frac{1}{2\sqrt 2} i
\gamma_5\gamma_\mu (1+\gamma_\nu\frac{\partial_\nu}{\sqrt{\partial^2}}) 
A_\mu^q(X) ,
\end{eqnarray}    
where $S^q(X)$ and $A_\mu^q(X)$ represents the local pseudoscalar
and vector meson field operators of $D$-type, respectively.\\
For extremely relativistic-type spinors
\begin{eqnarray}
\frac{1}{2\sqrt 2} &i& 
\gamma_5 (1-\gamma_\nu\frac{\partial_\nu}{\sqrt{\partial^2}}) P_s^{(ER)}(X), \ \ \ 
\frac{1}{2\sqrt 2} i 
\gamma_\mu (1-\gamma_\nu\frac{\partial_\nu}{\sqrt{\partial^2}}) V_\mu^{(ER)}(X) ,
\end{eqnarray}    
where $P_s^{(ER)}(X)$ and $V_\mu^{(ER)}(X)$ represents the local pseudoscalar
and vector meson field operators of extremely relativistic-type, respectively.

These local meson field representation of the composite quark spinor WF
guarantees the crossing symmetry of the interaction 
among composite mesons, which are induced from 
the crossing symmetric interaction between constituent quarks and antiquarks.

\section{orthonormality relation for spinor WF}
\subsection{Bargmann-Wigner (BW) bases}
The Pauli-conjugate of the annihilation parts $W_i^{(+)}$ of WF, 
$\bar W_i^{(-)}(\equiv \gamma_4 W_i^{(+)\dagger } \gamma_4 )$, 
are related with the creation parts as
\begin{eqnarray}
(\bar W_i^{(-)}) &=& (\bar U_{P_s}^{(-)},\bar U_{V_\mu}^{(-)},\bar C_S^{(-)},\bar C_{A_\mu}^{(-)},
\bar D_S^{(-)},\bar D_{A_\mu}^{(-)}, \bar V_{P_s}^{(-)},\bar V_{V_\mu}^{(-)}) \nonumber\\
 &=& (U_{P_s}^{(-)},U_{V_\mu}^{(-)},D_S^{(-)},D_{A_\mu}^{(-)},
C_S^{(-)},C_{A_\mu}^{(-)}, V_{P_s}^{(-)},V_{V_\mu}^{(-)} ).
\end{eqnarray}
They are orthonormal to the annihilation WF $W_i^{(+)}$ as 
\begin{eqnarray}
\langle \bar W_i^{(-)}W_j^{(+)} \rangle  &=& \epsilon_{i}\delta_{ij} \\
 (\epsilon_{P_s^{(NR)}},\epsilon_{V_\mu^{(NR)}},\epsilon_{S^{(\bar q)}},&&
\epsilon_{A_\mu^{(\bar q)}},\epsilon_{S^{(q)}},\epsilon_{A_\mu^{(q)}},
\epsilon_{P_s^{(ER)}},\epsilon_{V_\mu^{(ER)}}) 
= (-1,-1,1,1,1,1,-1,-1) .\nonumber
\end{eqnarray}
By using this orthonormality relation
we can decompose the general spinor WF $\Psi^{(+)}$ into the meson
components,  
$(\ \  \phi_i^{(+)} \ \ )=(P_s^{(+)(NR,ER)}$, $S^{(+)(\bar q,q)}$, 
$V_\mu^{(+)(NR,ER)}$, $A_\mu^{(+)(\bar q,q)} )$.
\begin{eqnarray}
 \Psi^{(+)} &=& \sum_{i} \   W_i^{(+)}\  \phi_i^{(+)}  .\ \ \   
 \phi_i^{(+)} = \epsilon_i \langle \bar W_i^{(-)} \Psi^{(+)}   \rangle
\end{eqnarray}

\subsection{chiral bases---light quark $q\bar q$-meson systems}
For description of the light quark $q\bar q$-meson systems the chiral $(N,E)$ bases 
of spinor WF are expected to be more effective. 
They are obtained by the linear combination of 
BW bases as explained in the text.

The annihilation WF $W_i^{(+)}$ and creation WF $W_i^{(-)}$ are given, respectively, by
\begin{eqnarray}
(\  W_i^{(+)}(P)\ ) &=& (U_{P_s}^{(+)(N)}, C_{S}^{(+)(N)},  U_{V_\mu}^{(+)(N)},  
C_{A_\mu}^{(+)(N)};  
         U_{P_s}^{(+)(E)}, C_{S}^{(+)(N)}, U_{V_\mu}^{(+)(E)},  C_{A_\mu}^{(+)(E)} ) \nonumber\\
&=& \frac{1}{2}(i\gamma_5,1,i\tilde\gamma_\mu , i\gamma_5\tilde\gamma_\mu ;
-\gamma_5v\cdot\gamma ,-v\cdot\gamma ,-i\sigma_{\mu\nu}v_\nu ,
\gamma_5 \sigma_{\mu\nu}v_\nu ) \nonumber\\
(\  W_i^{(-)}(P)\ ) &=& (U_{P_s}^{(-)(N)}, C_{S}^{(-)(N)},  U_{V_\mu}^{(-)(N)},  C_{A_\mu}^{(-)(N)};  
         U_{P_s}^{(-)(E)}, C_{S}^{(-)(N)}, U_{V_\mu}^{(-)(E)},  C_{A_\mu}^{(-)(E)} ) \nonumber\\
&=& \frac{1}{2}(i\gamma_5,1,i\tilde\gamma_\mu , i\gamma_5\tilde\gamma_\mu ;
\gamma_5v\cdot\gamma ,v\cdot\gamma ,i\sigma_{\mu\nu}v_\nu ,
-\gamma_5 \sigma_{\mu\nu}v_\nu ) ,
\end{eqnarray}
which are related through crossing rules of constituent spinors with each other.
 
The Pauli-conjugate of the creation spinor 
$\bar W_i^{(-)}(\equiv \gamma_4 W_i^{(+)\dagger}   \gamma_4)$  are equal to 
the annihilation $W_i^{(-)}$.
\begin{eqnarray}
\bar  W_i^{(-)}(P) & (\equiv & \gamma_4 W_i^{(+)\dagger}   \gamma_4)
=W_i^{(-)}(P) \ \ \ \ \nonumber\\
 &{\rm for}&\ \  \phi_i=P_s^{(N)},S^{(N)},V_\mu^{(N)},A_\mu^{(N)};
P_s^{(E)},S^{(N)},V_\mu^{(E)},A_\mu^{(E)}  .
\end{eqnarray}  
They satisfy orthonormality relation:
\begin{eqnarray}
\langle \bar W_i^{(-)} W_j^{(+)} \rangle &=& \epsilon_i \delta_{ij}\nonumber\\
(\  \epsilon_i\ ) &=& (\epsilon_{P_s^{(N)}},\epsilon_{S^{(N)}},\epsilon_{V_\mu^{(N)}},
\epsilon_{A_\mu^{(N)}};\epsilon_{P_s^{(E)}},\epsilon_{S^{(N)}},
\epsilon_{V_\mu^{(E)}},\epsilon_{A_\mu^{(E)}}  )\nonumber\\
&=& (-1,1,-1,1;1,-1,-1,1)
\end{eqnarray}

By using this orthonormality relation the general WF $\Psi^{(+)}$ are 
decomposed 
into the meson components $(\ \  \phi_i^{(+)} \ \ )=(P_s^{(+)(N,E)}$, $S^{(+)(N,E)}$, 
$V_\mu^{(+)(N,E)}$, $A_\mu^{(+)(N,E)} )$.
\begin{eqnarray}
 \Psi^{(+)} &=& \sum_{i} \   W_i^{(+)}\  \phi_i^{(+)}  .\ \ \   
 \phi_i^{(+)} = \epsilon_i \langle \bar W_i^{(-)} \Psi^{(+)}   \rangle
\end{eqnarray}

\section{Chiral transformation for spinor WF of light quark $q\bar q$ mesons}
$SU(3)$ chiral transformation for the annihilation part of spinor WF 
of light quark $q\bar q$ mesons is given by 
\begin{eqnarray}
\Psi_A{}^{(+)B}(P,x) &\rightarrow & 
   [e^{i\frac{\alpha^i\lambda^i}{2}\gamma_5}\Psi^{(+)} (P,x)
    e^{i\frac{\alpha^i\lambda^i}{2}\gamma_5}]_A{}^B .
\label{eqB1}
\end{eqnarray}
For the infinitesimal transformation 
$\{ i\frac{\alpha^i\lambda^i}{2}\gamma_5,\Psi^{(+)}(P,x)\}$
the respective meson spinor WF are transformed as
\begin{eqnarray} 
P_s^{(N)}\cdot U_{P_s}^{(N)}  &\rightarrow&
\{ \frac{\alpha^i\lambda^i}{2},P_s^{(N)} \}\cdot \frac{1}{2}
 \{ i\gamma_5, U_{P_s}^{(N)} \} = d^{ijk}\ \alpha^i P_s^{(N)j}
       \frac{\lambda^k}{2} \cdot  C_S^{(N)}  \nonumber \\
S^{(N)}\cdot C_{S}^{(N)}  &\rightarrow&
\{ \frac{\alpha^i\lambda^i}{2},S^{(N)} \}\cdot \frac{1}{2}
 \{ i\gamma_5, C_{S}^{(N)} \} = d^{ijk}\ \alpha^i S^{(N)j}
       \frac{\lambda^k}{2} \cdot  U_{P_s}^{(N)}  \nonumber \\
V_\mu^{(N)}\cdot U_{V_\mu}^{(N)}  &\rightarrow&
[ \frac{\alpha^i\lambda^i}{2},V_\mu^{(N)} ] \cdot \frac{1}{2}
 [ i\gamma_5, U_{V_\mu}^{(N)} ] = - f^{ijk}\ \alpha^i V_\mu^{(N)j}
       \frac{\lambda^k}{2} \cdot  C_{A_\mu}^{(N)}  \nonumber \\
A_\mu^{(N)}\cdot C_{A_\mu}^{(N)}  &\rightarrow&
[ \frac{\alpha^i\lambda^i}{2},A_\mu^{(N)} ] \cdot \frac{1}{2}
 [ i\gamma_5, C_{A_\mu}^{(N)} ] = - f^{ijk}\ \alpha^i A_\mu^{(N)j}
       \frac{\lambda^k}{2} \cdot U_{V_\mu}^{(N)}  \nonumber \\
P_s^{(E)}\cdot U_{P_s}^{(E)}  &\rightarrow&
[ \frac{\alpha^i\lambda^i}{2},P_s^{(E)} ] \cdot \frac{1}{2}
 [ i\gamma_5, U_{P_s}^{(E)} ] = - f^{ijk}\ \alpha^i P_s^{(E)j}
       \frac{\lambda^k}{2} \cdot C_{S}^{(E)}  \nonumber \\
S^{(E)}\cdot C_{S}^{(E)}  &\rightarrow&
[ \frac{\alpha^i\lambda^i}{2},S^{(E)} ] \cdot \frac{1}{2}
 [ i\gamma_5, C_{S}^{(E)} ] = - f^{ijk}\ \alpha^i S^{(E)j}
       \frac{\lambda^k}{2} \cdot U_{P_s}^{(E)}  \nonumber \\
V_\mu^{(E)}\cdot U_{V_\mu}^{(E)}  &\rightarrow&
\{ \frac{\alpha^i\lambda^i}{2},V_\mu^{(E)} \}\cdot \frac{1}{2}
 \{ i\gamma_5, U_{V_\mu}^{(E)} \} = d^{ijk}\ \alpha^i V_\mu^{(E)j}
       \frac{\lambda^k}{2} \cdot  C_{A_\mu}^{(E)}  \nonumber \\
A_\mu^{(E)}\cdot C_{A_\mu}^{(E)}  &\rightarrow&
\{ \frac{\alpha^i\lambda^i}{2},A_\mu^{(E)} \}\cdot \frac{1}{2}
 \{ i\gamma_5, C_{A_\mu}^{(E)} \} = d^{ijk}\ \alpha^i A_\mu^{(E)j}
       \frac{\lambda^k}{2} \cdot  U_{V_\mu}^{(E)}\ \ . 
\end{eqnarray}
For the finite $U(1)$ chiral transformation 
\begin{eqnarray}
\Psi_A{}^{(+)B}(P,x) &\rightarrow & 
   [e^{i\alpha\gamma_5}\Psi^{(+)} (P,x)
    e^{i\alpha\gamma_5}]_A{}^B ,
\end{eqnarray}
the spinor WF for $V_\mu^{(N)},\ A_\mu^{(N)},\ P_s^{(E)}$ and $S^{(E)}$ are
invariant, while
\begin{eqnarray}
P_s^{(N)}\cdot U_{P_s}^{(N)}  &\rightarrow&
   P_s^{(N)} \cdot ( U_{P_s}^{(N)}\ {\rm cos}\ 2\alpha 
   - C_s^{(N)}\ {\rm sin}\ 2\alpha ) ,\nonumber \\
S^{(N)}\cdot C_{S}^{(N)}  &\rightarrow&
   S^{(N)} \cdot ( C_s^{(N)}\ {\rm cos}\ 2\alpha 
   + U_{P_s}^{(N)}\ {\rm sin}\ 2\alpha ) , \nonumber \\
V_\mu^{(E)}\cdot U_{V_\mu}^{(E)}  &\rightarrow&
   V_\mu^{(E)} \cdot ( U_{V_\mu}^{(E)}\ {\rm cos}\ 2\alpha 
   + C_{A_\mu}^{(E)}\ {\rm sin}\ 2\alpha ) , \nonumber \\
A_\mu{(E)}\cdot C_{A_\mu}^{(E)} & \rightarrow &
   A_\mu^{(E)} \cdot ( C_{A_\mu}^{(E)}\ {\rm cos}\ 2\alpha 
   - U_{V_\mu}^{(E)}\ {\rm sin}\ 2\alpha ) .
\end{eqnarray}


\begin{thebibliography}{9}
\bibitem{rf0} S. Ishida and M. Ishida, proc. of international workshop on 
$e^+ e^-$ Collisions from $\phi$ to $J / \Psi$, 
Novosibirsk, Russia, 1-5 Mar 1999, hep-ph/9905258 .\\
S. Ishida, M. Ishida and T. Maeda, proc. of 8th International 
Conference on Hadron Spectroscopy (HADRON 99), 
Beijing, China, 24-28 Aug 1999, hep-ph/0002095.
\bibitem{rf1}S. Ishida, M. Y. Ishida and M. Oda, 
   {\it Prog. Theor. Phys.} {\bf 93}, 939 (1995). 
\bibitem{rf3}S. Ishida,
    {\it Prog. Theor. Phys.} {\bf 46}, 1570 and 1950 (1971). 
\bibitem{rf2}H. Yukawa,
   {\it Phys. Rev.} {\bf 91}, 415 and 416 (1953). 
\bibitem{rfmass}R. Oda, K. Yamada, S. Ishida, M. Sekiguchi and H. Wada,
Prog. Theor. Phys. {\bf 102}(1999), 297.
\bibitem{rf5} S. Ishida, A. Morikawa and M. Oda,
Prog. Theor. Phys. {\bf 99}(1998), 257.\\
M. Oda, M. Y. Ishida and S. Ishida, Nihon Univ. Preprint
NUP-A-94-7.\\
S. Ishida, M. Y. Ishida and M. Oda, Prog.
Theor. Phys. {\bf 93}(1995), 781.\\
M. Y. Ishida, S. Ishida and M. Oda, Prog.
Theor. Phys. {\bf 98}(1997), 159.\\
M. Oda, M. Y. Ishida and S. Ishida, Prog.
Theor. Phys. {\bf 101}(1999), 1285.\\
R. Mohanta, A. K. Giri, M. P. Khanna,
M. Y. Ishida, S. Ishida and M. Oda, Prog.
Theor. Phys. {\bf 101}(1999), 947, 959, 1083; {\bf 102}(1999), 645.
\bibitem{rfsca} M. D. Scadron, Phys. Rev. {\bf D26} (1982), 239; hep-ph/9710317.\\
R. Delbourgo and M. D. Scadron, Phys. Rev. Lett. {\bf 48} (1982), 379.\\
V. Elias and M. D. Scadron, Phys. Rev. Lett. {\bf 53} (1984), 1129; Mod. Phys. Lett.
{\bf A10} (1995), 251.
\bibitem{rfRupp} E. Beveren, T. A. Rijken, K. Metzger, C. Dullemond, G. Rupp
and J. E. Ribeiro, Z. Phys. {\bf C30} (1986), 615.
\bibitem{rftorn} N. A. Tornqvist Phys.Lett.{\bf B426} (1998), 105. 
\bibitem{rfmy} M. Ishida, Prog. Theor. Phys. {\bf 101} (1999), 661.
\bibitem{rfpipi1} S. Ishida, M. Ishida, H. Takahashi, T. Ishida, K. Takamatsu
and T. Tsuru, Prog. Theor. Phys. {\bf 95} (1996), 745; {\bf 98} (1997), 1005.
\bibitem{rfpipi2} N. N. Achasov and G. N. Schestakov,  Phys. Rev. {\bf D49} (1994), 5779.
\bibitem{rfpipi3} R. Kaminski, L. Lesniak and J. -P. Maillet, 
Phys. Rev. {\bf D50} (1994), 3145.
\bibitem{rfpipi4} N. A. Tornqvist, Z. Phys. {\bf C68} (1995), 647.\\
N. A. Tornqvist and M. Roos, Phys. Rev. Lett. {\bf 76} (1996), 1575.
\bibitem{rfpipi5} M. Harada, F. Sannino and J. Schechter, Phys. Rev. {\bf D54} (1996), 1991.
\bibitem{rfprod1} T. Ishida, Doctor Thesis, University of Tokyo (1996);
KEK Report 97-8 (1997).
\bibitem{rfprod2} K. Takamatsu,
M. Ishida, S. Ishida, T. Ishida and T. Tsuru, hep-ph/9712232,  
in {\em Proc.\ of\ Int.\ Conf.\ on\ Hadron\ 97,\ 
Brookhaven\ National\ Laboratory,} ed. S. U. Chung and H. J. Willutzki,
{\em AIP\ conference\ proceedings\ 432,\  Upton,\ NY,\ 1997.}
\bibitem{rfprod3} D. Alde et al., Phys. Lett. {\bf B397} (1997), 350.
\bibitem{rfprod4} M. Ishida, T. Komada, S. Ishida, T. Ishida, 
K. Takamatsu, T. Tsuru,  
in proc. of 8th International Conference on Hadron Spectroscopy (HADRON 99), 
Beijing, China, 24-28 Aug 1999: NUP-A-99-17, hep-ph/9910248 .
\bibitem{rfkappa} S. Ishida, M. Ishida, T. Ishida, K. Takamatsu
and T. Tsuru, Prog. Theor. Phys. {\bf 98} (1997), 621.
\bibitem{rfsche} D. Black, A. H. Fariborz, F. Sannino and J. Schechter, 
Phys. Rev. {\bf D58} (1998), 054012. 
\bibitem{rfeta1}S. Fukui et al., Phys. Lett. {\bf B267} (1991), 293.
\bibitem{rfeta2}D. Alde et al, Phys. Atomic Nuclei, {\bf 60} (1997), 386.\\
C. Caso et al., Euro. Phys. J. {\bf C3} (1998), 386.  
\bibitem{rfeta3}A. Ando et al., Phys. Rev. Lett. {\bf 57} (1986), 1296.
\bibitem{rfeta4}M. G. Rath et al., Phys. Rev. {\bf D40} (1989), 693.\\
Z. Bai et al., Phys. Rev. Lett. {\bf 65} (1990), 2507.\\
C. Caso et al., Euro. Phys. J. {\bf C3} (1998), 398
\bibitem{rf6}D. Alde et al., Phys. Lett. {\bf B205}(1988), 397.\\
H. Aoyagi et al.,  Phys. Lett. {\bf B314}(1993), 246.\\
S. U. Chung et al., Phys. Rev. {\bf D60}(1999), 092001.
\bibitem{rfchung}S. U. Chung et al.,in proc. of 8th International 
Conference on Hadron Spectroscopy (HADRON 99), Beijing, China, 24-28 Aug 1999.
\bibitem{rfR1}C. Caso et al., Euro. Phys. J. {\bf 3} (1998), 380.
\bibitem{rfR2}S. Ishida, M. Oda, H. Sawazaki and K. Yamada, Prog. Theor. Phys. {\bf 88}
(1992), 89.
\bibitem{rfwaka}A. Wakabayashi, master thesis, Nihon University, 2000.
\bibitem{rfkoba} M. Kobayashi, in proc. of workshop on {\em Possible\ Existence\ of\ 
$\sigma$-meson\ and\ its\ Implications\ to\ hadron\ physics}, at YITP in Kyoto, June 12-14, 2000.
to be published in Soryuusiron Kenkyuu, KEK-Report.
\bibitem{rfR3} T. Maeda,  master thesis, Nihon University, 2000.
\bibitem{rfR4} S. Goldfarb, in Frascati Physics Series Vol. XV, 321(1999).
Proc. of the Workshop on Hadron Spectroscopy, March 8-12, 1999.
\bibitem{rf31} S. Ishida and P. Roman, Phys. Rev. {\bf 172} (1968), 1684. 
\end{thebibliography}
\end{document}